# High-resolution label-free imaging of tissue morphology with confocal phase microscopy

Martin Schnell[1*], Shravan Gupta[2], Tomasz P. Wrobel[1,3], Michael G. Drage[4], Rohit Bhargava[1,5], P. Scott Carney[2]

1  Beckman Institute for Advanced Science and Technology, University of Illinois at Urbana-Champaign, Urbana, IL 61801, USA
2  The Institute of Optics, University of Rochester, 480 Intercampus Drive, Rochester, NY 14620, USA
3  Solaris National Synchrotron Radiation Centre, Jagiellonian University, Czerwone Maki 98, 30-392 Krakow, Poland
4  Department of Pathology and Laboratory Medicine, University of Rochester Medical Center, Rochester, NY  14642, USA
5  Department of Bioengineering, Department of Chemical and Biomolecular Engineering, Cancer Center at Illinois, Department of Electrical and Computer Engineering, and Department of Chemistry, University of Illinois at Urbana-Champaign, Urbana, IL 61801, USA
* schnelloptics@gmail.com

**Abstract (140 words)**
Label-free imaging approaches seek to simplify and augment histopathologic assessment by replacing the current practice of staining by dyes to visualize tissue morphology with quantitative optical measurements. Quantitative phase imaging (QPI) operates with visible/UV light and thus provides a resolution matched to current practice. Here we introduce and demonstrate confocal QPI for label-free imaging of tissue sections and assess its utility for manual histopathologic inspection. Imaging cancerous and normal adjacent human breast and prostate, we show that tissue structural organization can be resolved with high spatial detail comparable to conventional H&E stains. Our confocal QPI images are found to be free of halo, solving this common problem in QPI. We further describe and apply a virtual imaging system based on Finite-Difference Time-Domain (FDTD) calculations to quantitatively compare confocal with wide-field QPI methods and explore performance limits using numerical tissue phantoms.

**Key Words:** quantitative phase imaging, confocal microscopy, label free imaging, interferometry, pathology, cancer diagnosis

1. Introduction

The microscopic examination of stained tissue for structural and functional changes is the standard method for detecting and grading most forms of human cancer [1]. For example, the hematoxylin and eosin (H&E) stain allows a trained observer to differentiate epithelial cells from the surrounding stroma. Morphological features of epithelial cells and their organization, further, are the basis for cancer diagnoses and studying progression. Scientific advances in the fields of microscopy could lead to augmenting and automation of histopathologic studies by combining label-free, quantitative imaging approaches with machine learning [2,3]. These techniques can provide the consistency in measurement of optical properties of tissue that cannot be manually assessed, nor quantified by using stains. Together, this can obviate the need for staining, allow for objective decision making based on pattern recognition and can challenge current histopathologic practice.

Quantitative Phase Imaging (QPI) is a particularly promising label-free method. It combines wide-field (WF) optical microscopy with interferometry to reveal the structure of unlabeled, transparent samples [4–14]. The underlying principle is that spatial variations in the refractive index introduce changes in the relative phase of a light wave as it passes through the sample. QPI measures the phase of the transmitted light quantitatively and can provide a viable alternative to established label-based imaging methods such as fluorescence imaging of cells. Applied to tissue, QPI has shown to reveal the microstructural organization without staining and at the resolution of optical microscopy. Demonstrations of QPI-based histopathologic investigations [2] include prostate [15,16], colon [17,18], breast [19,20]. The label-free approach promises a simpler pathology work flow by removing the need for sample staining while providing quantitative morphologic parameters to detect and grade cancer [2,16,17].

Although QPI data shows correspondence to reference images of H&E-stained tissue, a detailed comparison of image quality and optimization of image contrast to mimic and enhance that of stained tissue images is still widely unaddressed. This might be understood by considering that QPI faces several fundamental and technical challenges, which require a compromise to be made between image quality, usability and speed [14]. For example, some QPI implementations are affected by mechanical vibrations and air density fluctuations that cause temporal path length variations and lead to errors in the phase measurement. Hence, design of QPI systems and their operation in a suitable environment may not be compatible with the rugged environment of a pathology laboratory. To reduce stability requirements, common-path QPI methods were developed that largely cancel the effect of these temporal instabilities on the phase measurement. However, common-path methods are challenged by halo and shade-off appearing in the phase images, which are object-dependent effects that appear as negative phase contrast at edges of objects and reduced contrast in the center of large area objects [21–23]. Further challenges were found with shot noise and speckle, which could be addressed by using high full well camera technology [24] and low coherence light sources [25–27], respectively.

In the search for highest image quality, speed and ease of use, confocal microscopy has received less attention that its WF counterparts. This is surprising because confocal microscopy is a widely used method in biomedical imaging. With fluorescence contrast, for example, it provides an unobscured view on the details of biological structure owing to the rejection of out-of-focus light provided by spatial pinhole filtering. When combined with QPI, it offers halo- and speckle free phase imaging with standard monochromatic laser sources. Interferometric confocal microscopy has been employed for static [28–30] and dynamic [31–33] optical metrology and later to label-free imaging of cells [34–36]. Optical coherence tomography has been applied to phase imaging of single cells and depth-resolved phase imaging of tissue [37,38]. Recently, synthetic optical holography (SOH) [39] was demonstrated as a holographic approach to confocal QPI [40–42]. Implemented in a commercial confocal instrument, SOH allowed confocal QPI based on a beam scanning approach with frame acquisition times on the order of seconds, thus mitigating the slow scanning speed of initial demonstrations of confocal QPI based on sample scanning [43,44]. All these advances indicate that confocal QPI has become a promising candidate for practical, rapid and high-resolution label-free imaging of biological specimen.

Here, we demonstrate confocal QPI of tissue sections as a means to provide label-free, high resolution morphology images close to the level of detail of H&E. We image malignant and normal-adjacent tissue sections of breast, prostate, stomach and cerebrum and subsequently compare spatial detail and contrast of tissue morphology with gold-standard H&E images obtained from adjacent tissue sections. To compare image contrast with established QPI methods, we apply a virtual imaging system based on Finite-Difference Time-Domain (FDTD) and synthesize QPI images for relevant test samples. This development also shows how refractive index models of tissue sections can serve as a testbed for predicting performance of future QPI instruments.

## 2. Method

For our study, multi-organ tumor tissue microarrays (TMA) were purchased from US Biomax Inc. (Serial# MC246b). The TMA contained multiple organ tumor and matched adjacent tissue microarray of 12 types of organs. Each core measured 1.5 mm in diameter and 5 μm in thickness. For phase imaging, an unstained tissue was ordered on a reflective glass slide (Deposition Research Lab, Inc., St. Charles, Missouri), deparaffinized in xylene and subsequently coverslipped using Permount (Fisher Scientific) as mounting medium. For comparison, a consecutive H&E-stained tissue section was obtained as well.

Figure 1a illustrates quantitative confocal phase imaging of the TMA. We applied sinusoidal-phase synthetic optical holography, a recently introduced holographic modality of confocal phase imaging [43,44]. Briefly, the sample is imaged in transflection geometry with the confocal focus placed at the reflective layer of the glass slide, using a commercial confocal microscope (Nikon, model: A1R). Illumination wavelength was 561nm at 2.3% power setting. At each position $(x, y)$ on the sample, the local refractive index imparts a specific phase shift on the reflected beam that carries information on the local tissue structure. A Mirau interference objective (Nikon, model: CF IC EPI Plan DI 20×, 0.4 NA) detects this phase shift by interfering the reflected beam from the sample, $U_S$, with an internally generated reference beam, $U_R$. While the focus is raster-scanned across the sample, the sample is vertically vibrated at sub-micrometer amplitude with a piezo nanopositioner stage (Mad City Labs Inc., model: Nano-Z100). This introduces a sinusoidal-wave phase modulation across the image to encode amplitude and phase information of the sample in cross-pixel information, that is, holographically. The combined sample and reference beams are detected with the photomultiplier tube of the confocal microscope. Since the combined beam is of the same wavelength as the laser emission, an 80:20 beam splitter is inserted in lieu of a dichroic beam splitter, all filters are removed in front of the detector and the pinhole is fully opened. Figure 1b shows an example confocal hologram image as obtained with the TMA, exhibiting a finely spaced fringe pattern (Fig. 1c). The sinusoidal phase modulation yields a characteristic pattern of direct and conjugate terms in the Fourier transform of the hologram (Fig. 1d), each pertaining to an individual plane-wave component of the sinusoidal phase modulation. By combining terms 1 and 2, amplitude and phase images of the sample could be retrieved. The amplitude image (Fig. 1e) shows homogenous contrast with the tissue structure appearing only with faint contrast, which confirms that the coverslipped tissue section behaves mainly as a phase-contrast only object. The phase image (Fig. 1f) reveals rich information on the tissue structure, which forms the basis for label-free imaging.

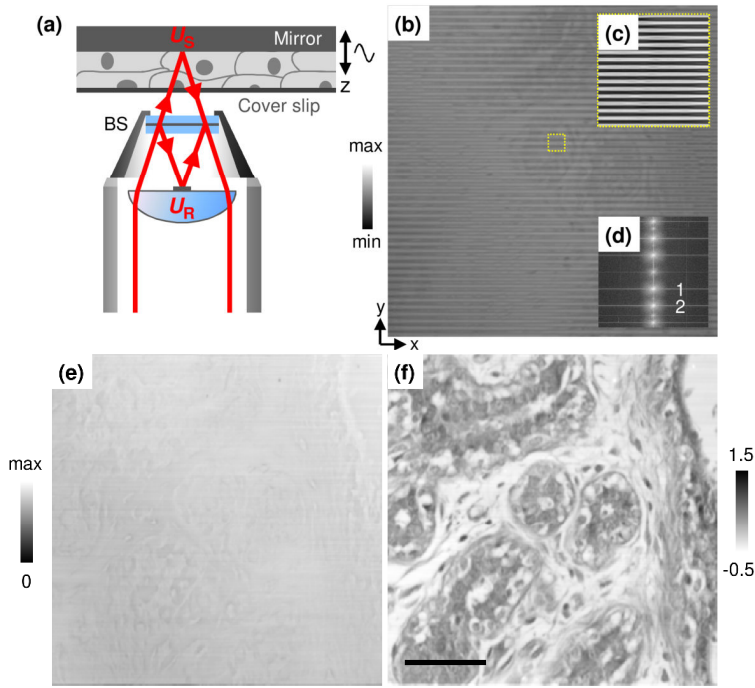

Fig. 1: Confocal phase imaging of tissue sections mounted on reflective glass slide with sinusoidal-wave synthetic optical holography. (a) Schematic (b) Example hologram and (c) digital zoom. (d) Fourier transform of the hologram. (e,f) Reconstructed amplitude and phase images.

Next, we assess the utility of confocal phase contrast to detect disease-induced morphologic changes in different organs. To this end, a set of confocal holograms was obtained by mosaic imaging of the entire tissue core (individual tiles covered 210 µm x 210 µm, 10% overlap). Phase images were reconstructed and subsequently stitched together with our own custom software. Phase drifts as a result of interferometer stabilities were removed by subtraction of horizontal line averages in each tile. Contrast between tiles was equalized by applying limited high-pass filtering to each tile (low frequency cut-off was 1/2 of the tile width). The stitched phase image of each tissue core was evaluated to assess the capability of confocal phase microscopy to spatially resolve tissue architecture and to explore how well the contrast of the common H&E stain was reproduced. Digital zooms are shown to determine cell shape and size, particularly shape, size and orientation of cell nuclei. For each organ type, phase images of benign and normal adjacent tissue cores are compared to reveal how well disease-induced changes in the tissue morphology can be recognized on both core and cellular level.

### 3. Experimental Results

For normal adjacent breast tissue, the overall tissue architecture was correctly reproduced (Fig. 2a). Benign epithelium appeared with a large phase shift (dark grey shade) with respect to the substrate, revealing the organization of breast lobules in the typically, well-organized form of clustered grapes. Stroma exhibited lower phase shift than epithelium (lighter grey) and adipocytes appeared as round disk of zero phase shift as a result of the removal of fatty tissue during sample preparation. A digital zoom (Fig. 2b) showed a rich level of detail of the tissue morphology. Individual lobules were well resolved. Stroma was nicely outlined, revealing a host of fibroblasts (F), blood vessels (B) and secretions (S). Loose stroma appeared with a lighter shading than dense stroma. Further digital zoom on an individual lobule (Fig. 2c) revealed that cells maintained their orientation. Nuclear features of epithelial cells were resolved well in shape; however, the contrast between nucleus and cytoplasm was weaker than in H&E. The edge of the basement membrane was clearly visible in some areas of the phase image, comparable to H&E. Digital zoom on stroma (Fig. 2d) resolved the spindly shape of fibroblasts with high resolution.

Confocal phase imaging of invasive ductal carcinoma (Fig. 2i) revealed a changed tissue morphology that was clearly different to the normal adjacent section in Fig. 2a. Digital zoom (Fig. 2j) showed fibrous stroma. Individual tumor nests were also present in the image, but difficult to discern because of similar shades of gray as the surrounding stroma at core-level view. Nevertheless, further zoom into the image (Figs 2k,l) revealed individual tumor cells recognizable by their outline. Loss of polarity could be recognized as well as enlarged nuclei of different sizes and with prominent, often multiple nucleoli (large phase shift, dark spots).

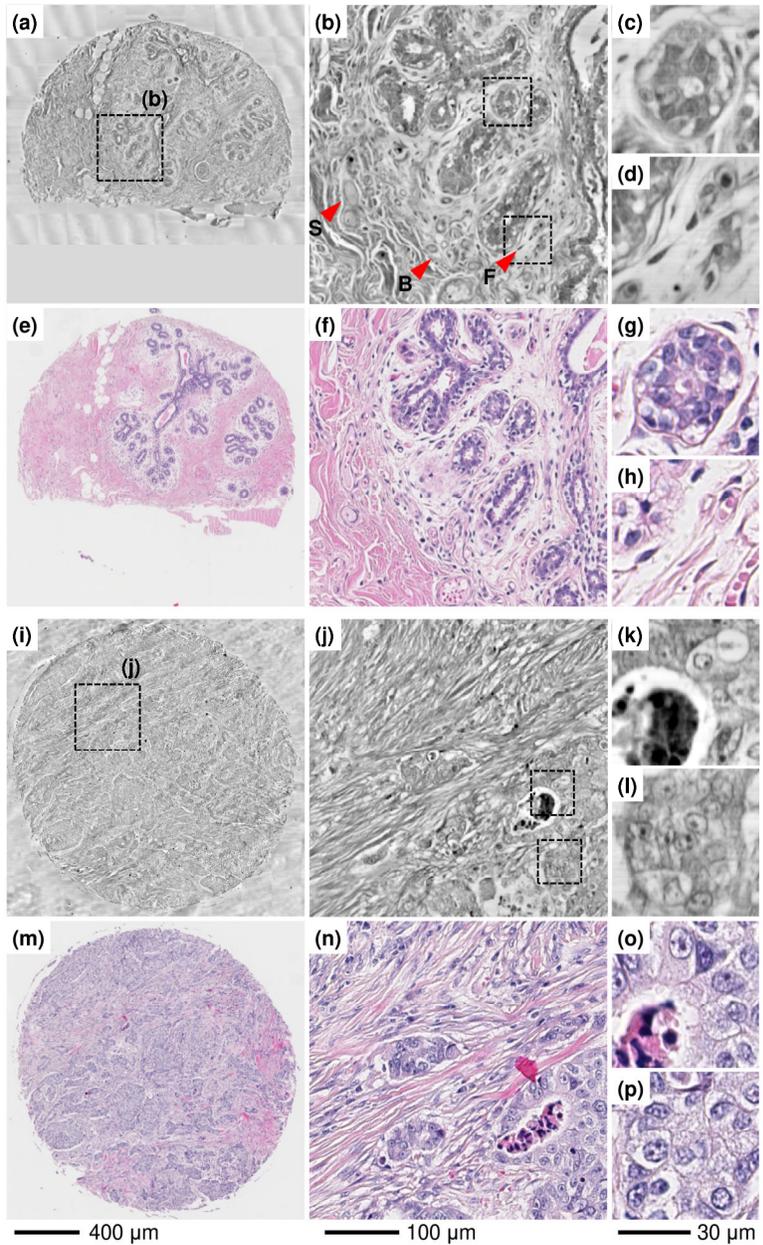

Fig. 2: Confocal phase images of normal adjacent breast (a-d) and histologic specimen corresponding to phase (e-h). Confocal phase imaging of invasive ductal carcinoma (i-l) and corresponding H&E image (m-p). Scale bar applies to entire column (core-level, medium zoom, high zoom setting).

With cancer-adjacent prostate tissue, ducts and acini were clearly revealed on a core-level view (Fig. 3a). Epithelial cell layers appeared slightly brighter than the surrounding stroma (arrowheads). Zoom on part of the duct revealed intraepithelial neoplasia seen as cell stacking (Fig. 3b). Cell polarization was clearly visible and basement membrane was outlined well. Cell nuclei appear darker than the surrounding cytoplasm. In comparison, prostate with adenocarcinoma lesions showed large central lumen filled with secretion and cell debris (Fig. 3e). The excess of glands in the surrounding tissue is evident when compared with the cancer-adjacent section. Stroma fibers and lymphocytes can be recognized by dark color. Several glands looked disorganized. Zoom on two individual glands revealed round and enlarged nuclei with prominent nucleoli (Fig. 3f). Generally, the contrast between cell nuclei and cytoplasm was lower compared to H&E, however, microscopic digital zoom on individual glands allowed to clearly discern nuclei owing the high spatial detailed provided by the phase images. Again, just as for breast tissue, the image quality allows an assessment of key features in both PIN and cancer.

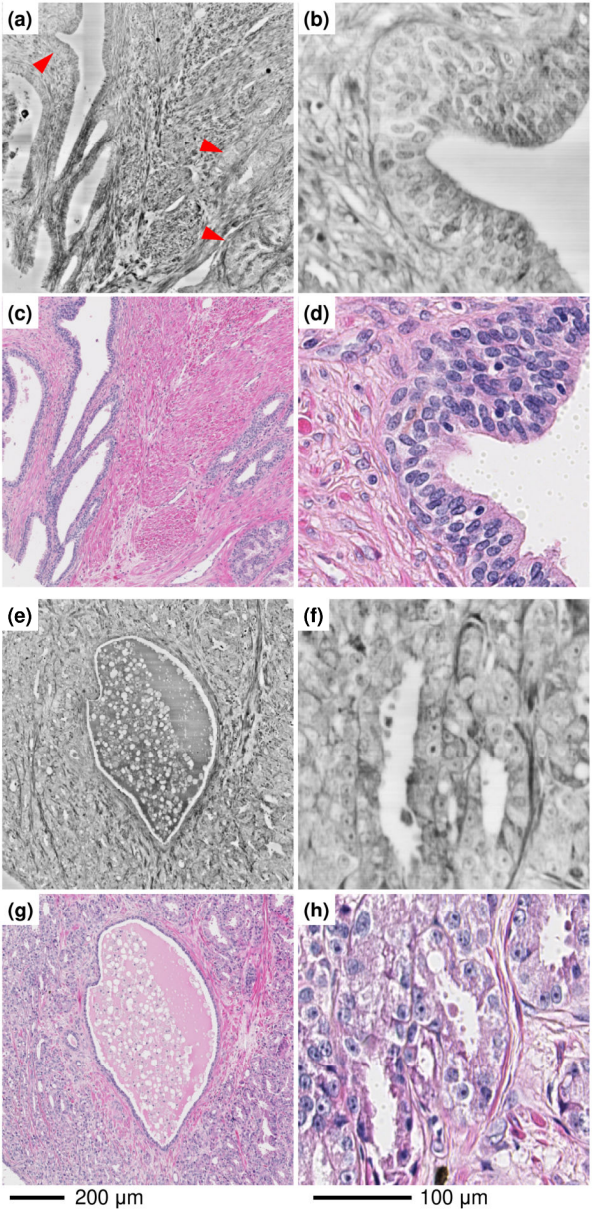

Fig. 3: Confocal phase images of cancer adjacent prostate (a-b) and corresponding H&E image (c-d). Confocal phase imaging of adenocarcinoma lesion in prostate (e-f) and corresponding H&E image (g-h).

While prostate and breast constitute the largest single cancers in men and women, we also evaluated phase contrast images for stomach and cerebrum. With adjacent normal stomach, the general organization of glands and cell polarization could be recognized (Fig. 4a). Adenocarcinoma lesions were revealed as large nests of tumor cells with enlarged nuclei and multiple prominent nucleoli (Fig. 4c). Lymphocyte infiltration in the stroma was observed. With cancer adjacent cerebrum, a sparse distribution of glial cells and individual pyramidal neurons was revealed (Fig. 4e). Glioblastoma lesions of cerebrum yielded a complex tissue architecture that was qualitatively correctly reproduced by phase imaging (Fig. 4g), as visual comparison with H&E demonstrates. These examples demonstrate the broad applicability of our method and utility for a variety of tissue morphologies.

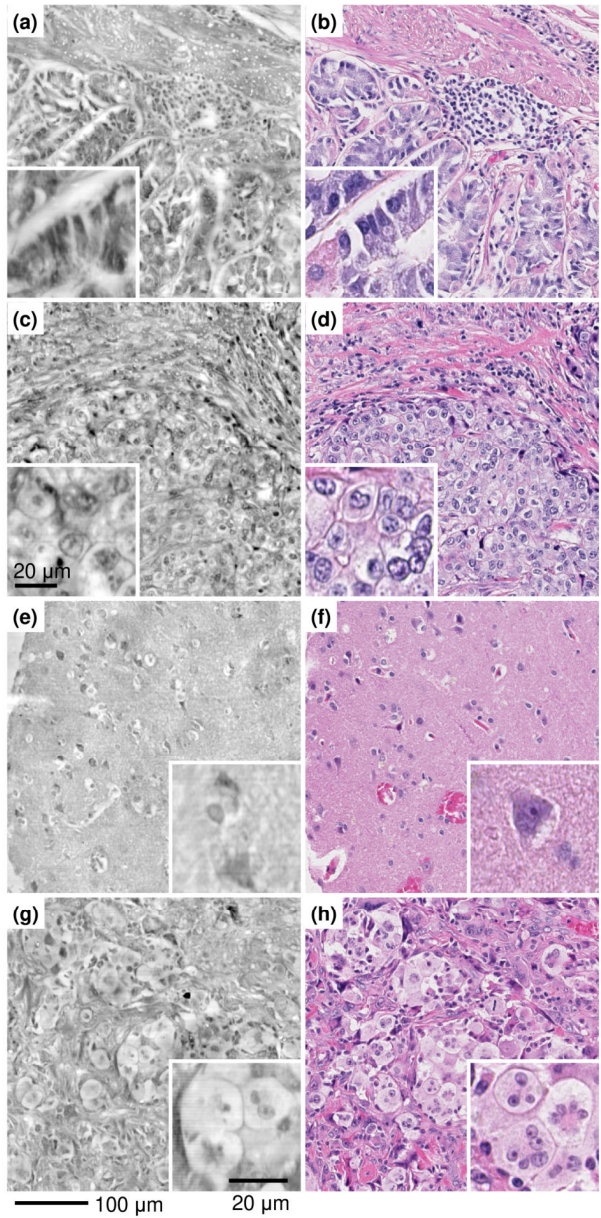

Fig. 4: Confocal phase images and H&E image of adjacent normal stomach (a-b), adenocarcinoma of stomach (c-d), cancer adjacent cerebrum (e-f), glioblastoma (g-h).

### 4. Virtual imaging system on a computer

The results above clearly show the potential of confocal phase imaging for stain-free imaging of tissue sections. We present below numerical calculations to obtain a prediction for the spatial resolution and image contrast that an optimized confocal system is capable of delivering. To this end, we applied the numerical algorithm described by Çapoğlu et al. [45] for implementing a virtual imaging system on a computer, which is based on rigorously solving Maxwell's equations using the finite-difference time domain (FDTD) method. We adopted this algorithm to model three relevant QPI methods: First, we assumed transmission-mode confocal QPI as a representative modality for the here presented method. Secondly, we considered traditional WF QPI where the reference field is a replica of the illumination field. Thirdly, we examined the widely employed method of common-path WF QPI where the reference field is derived by spatially filtering of the sample field. Using a commercial FDTD solver (Lumerical, Vancouver CA), we synthesized and quantitatively compared phase images for a set of test samples. In the following, we detail the specific adoptions made to the original algorithm in order to model QPI.

Figure 5(a) shows a schematic representation of a virtual imaging system of confocal microscopy with simulation segmented into scattering, collection and refocusing subsystems. Light scattering is calculated in the FDTD computational domain. The sample is illuminated with a Gaussian focus (NA = 0.8) that is injected by source S located at $z = -1$ µm, focused on the sample surface ($z = 0$) and polarized along the $x$-axis. The scattered field at the sample plane, $\mathbf{E}^{NF}(x,y)$, is detected with monitor M at $z = 1$ µm. A collection algorithm based on Fourier analysis produces the field at the far zone, $\mathbf{E}^{FF}(u_x, u_y)$, where $(u_x, u_y)$ are the corresponding diffraction orders. Diffraction orders $(u_x, u_y)$ falling outside the numerical aperture $\text{NA}_{obj} = 0.8$ of the collecting microscope objective are truncated. The refocusing algorithm Fourier-transforms the field passing the aperture and constructs the field distribution at the image plane, $\mathbf{E}^{IM}(x', y')$. Spatial filtering at the image plane with an infinitesimal pinhole is implemented by recording the refocused field at a single position located on the optical axis of the system, $\mathbf{E}^{IM}(0,0)$. To synthesize a confocal phase image, a set of simulations is run: While the sample is raster scanned in $(x, y)$ with a step size of 100 nm, the field $\mathbf{E}^{OBJ}(x,y) = \mathbf{E}^{IM}(0,0)\big|_{x,y}$ is computed for each sample position $(x, y)$. To generate the reference field, $\mathbf{E}^{REF}$, an empty sample is assumed, that is the structure is replaced by substrate material, and a single further simulation is run. Sample field, $\mathbf{E}^{OBJ}(x, y)$, and reference field, $\mathbf{E}^{REF}$, are interfered to obtain the confocal interference image,

$$I^{CF}_{\Delta\varphi}(x,y) = \sum_{n=x,y,z}\left|E_n^{OBJ}(x,y) + e^{i\Delta\varphi}E_n^{REF}\right|^2, \tag{1}$$

where index $n$ runs over the electric field components ($E_x$, $E_y$, $E_z$). The global phase between sample and reference field can be adjusted with $\Delta\varphi$.

Figure 5(b) shows the virtual imaging system used to model wide-field QPI. Bloch boundaries are assumed. The sample is illuminated with a single plane wave that is injected by source S located at $z = -1$ μm and at an angle as given by direction cosines, $(s_x, s_y)$. The scattered field from the sample, $\mathbf{E}^{NF}(x,y)$, is collected with monitor M at $z = 1$ μm and subsequently refocused at the image plane, yielding $\mathbf{E}^{OBJ} \equiv \mathbf{E}^{IM}(x,y)$. To implement Köhler illumination, it is necessary to run a series of simulations where the angle of the plane wave illumination is varied. More precisely, the direction cosines $(s_x, s_y)$ are assumed to be equally spaced and only those are considered that fall within the numerical aperture of the condenser $NA_{cond} = 0.09$, i.e. $s_x^2 + s_y^2 \leq NA^2$ as illustrated in Fig. 5c. This effort is needed because biological samples typically have key structural details comparable to the size of the wavelength and weak scattering may not be assumed (for details see ref. [45]). The spacing of the direction cosines, $\Delta s = 0.009 < 2\pi/kW$, is chosen to be small enough to avoid aliasing, where $k = 2\pi/\lambda$ is the wave vector of the laser wavelength and $W$ is the size of the FDTD box. Two simulations are run for each of the direction cosines: one simulation that considers the sample and a reference simulation, as described above. For common-path WF QPI, the reference field is obtained by spatial filtering of the far-zone collected light from the sample, $\mathbf{E}^{FF}(u_x, u_y)$, with a circular pinhole,

$$F^{PH}(u_x, u_y) = \begin{cases} 1 & \sqrt{u_x^2 + u_y^2} < 2\frac{2\pi}{kW} \\ 0 & \text{other} \end{cases}$$

The size of the pinhole was chosen to be 3 pixels rather than 1 pixel (in units of diffraction orders), which is still small but large enough correctly sample the field at the far zone, $\mathbf{E}^{FF}(u_x, u_y)$, as the direction of the plane wave illumination, $(s_x, s_y)$, is varied. For traditional WF QPI, the reference field, $\mathbf{E}^{REF}(x,y)$, is directly obtained from the reference simulation (empty sample). Sample field, $\mathbf{E}^{OBJ}(x,y)$, and reference field, $\mathbf{E}^{REF}(x,y)$, are interfered for each plane wave $(s_x, s_y)$ individually and the resulting interference images are added up incoherently to obtain the WF interference image,

$$I_{\Delta\varphi}^{WF}(x,y) = \sum_{s_x, s_y} \sum_{n=x,y,z} \left| E_{n,s_x,s_y}^{OBJ}(x,y) + e^{i\Delta\varphi} E_{n,s_x,s_y}^{REF}(x,y) \right|^2, \tag{2}$$

where the indices $(s_x, s_y)$ run over the direction cosines of the plane wave illumination and index $n$ runs over the electric field components $(E_x, E_y, E_z)$.

We retrieve the phase image of the sample, $\varphi(x,y)$, by applying 4-step phase shifting interferometry where we vary the global phase difference, $\Delta\varphi$, between the sample and reference field, $\mathbf{E}^{OBJ}(x,y)$ and $\mathbf{E}^{REF}(x,y)$, in steps of a quarter wavelength (0°, 90°, 180° and 270°) . The final phase image is obtained with the four-quadrant inverse tangent function,

$$\varphi = \text{atan2}(I_{90} - I_{270}, I_0 - I_{180}), \tag{3}$$

where indices $(x, y)$ are omitted for clarity. In this work, field calculations were done with built-in functions in Lumerical and phase interferometry was implemented in Matlab (MathWorks Inc, Natick MA). Importantly, the FDTD approach allows for accurately modelling of objects of arbitrary shape and with key structural details of the size of the wavelength, i.e. multiple scattering and resonance effects are taken into account. This capability is essential in modeling light scattering from microscopically heterogeneous samples such as tissue sections.

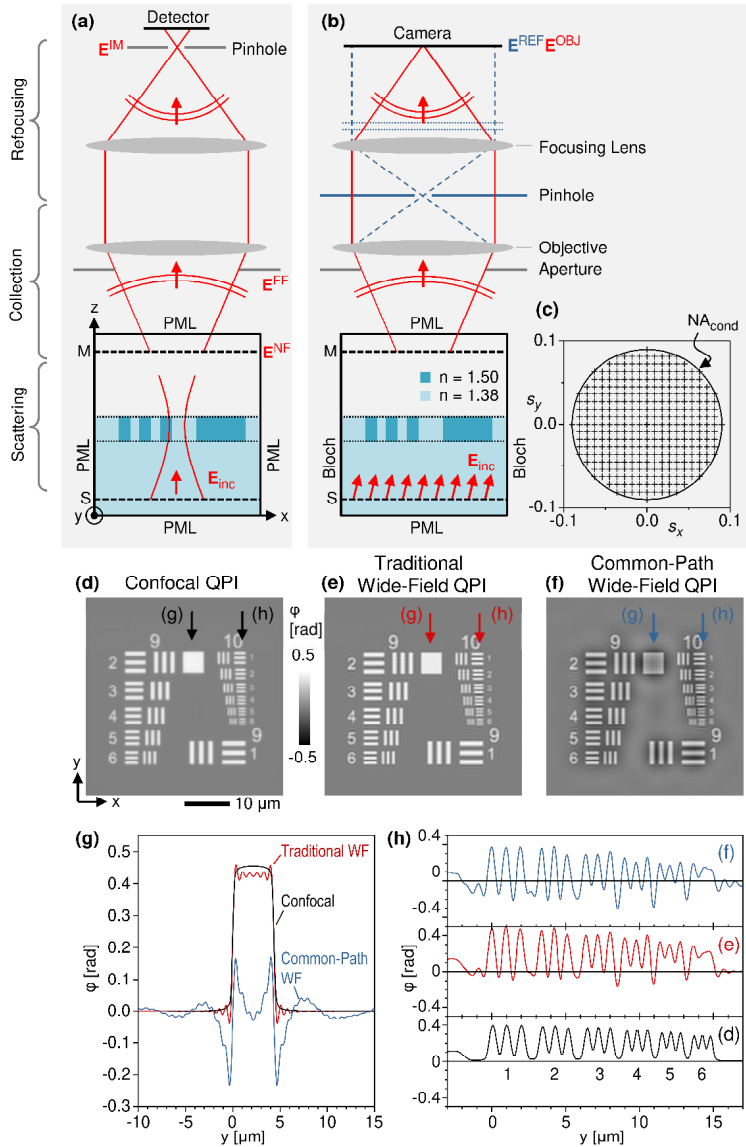

Fig. 5: Performance prediction of an idealized confocal QPI microscope and comparison to wide-field QPI methods. (a) Virtual imaging system based on FDTD describing transmission-mode confocal QPI. (b) Virtual wide-field QPI. PML: Perfectly matched layer boundary, Bloch: Bloch periodic boundary. (c) Equally-spaced plane waves describing Köhler illumination. (d-f) Synthesized quantitative phase images of a phase-object resolution test target at wavelength $\lambda = 561$ nm for confocal QPI ($\text{NA}_{\text{obj}} = 0.8$), traditional wide-field QPI and common-path wide-field QPI ($\text{NA}_{\text{cond}} = 0.09$ and $\text{NA}_{\text{obj}} = 0.8$). (g,h) Vertical line profiles across the square of group 9 and the bars of group 10, respectively, as indicated by the arrows in (d-f).

We first analyzed fundamental aspects of image contrast with a dielectric resolution test target based on group 9 and 10 of the USAF-1951 resolution test target. This test target contains numbers and bars with refractive index $n = 1.5$ that are embedded in a 250nm thick film of refractive index $n = 1.38$ and situated on an infinite substrate ($n = 1.38$). Comparison between the calculated phase images reveals the presence of halo and shade-off effects with common-path WF QPI (cf. Figs 5d-f). Halo is manifested as a dark ring around the structures, while shade-off appears as a reduced optical phase in the center of the structures, as it is further quantified by a line profile across the square of group 9 (Fig. 5g). Halo and shade-off are recognized as problems in QPI as they prevent accurate topography measurements and make interpretation of phase contrast in biological samples difficult [22,46,47]. In contrast, our calculations confirm that this problem is avoided with confocal and traditional WF QPI that are both free of halo and shade-off because the reference beam is generated directly from the illuminating beam.

Further differences can be observed with the presence of ringing artifacts at sharp edges (Fig. 5g) as well as a slight reduction in spatial resolution (Fig. 5h) with WF QPI as only elements 1-4 of group 10 are resolved according to the Rayleigh criterion. Interestingly, the line profile in Fig. 5h further shows significant negative phase contrast even in the case of traditional WF QPI. This is surprising as negative phase was typically attributed to common-path QPI only. Negative phase appears to be more strongly pronounced with the series of bars than it is with the isolated square (cf. Figs. 5g, 5h). In comparison, confocal QPI is free of ringing and negative phase contrast is not observed. Confocal QPI resolves all elements of group 10, corresponding to a spatial resolution of better than 1 µm ($\mathrm{NA_{OBJ}} = 0.8$, $\lambda = 561\,\mathrm{nm}$). We attribute these unexpected differences between confocal and WF QPI to the low numerical aperture of the condenser, in our calculation this is $NA_{\mathrm{cond}} = 0.09$. This choice follows a widely employed strategy to mitigate halo and shade-off by stepping down the condenser to establish sufficient spatial coherence at the sample [22,23]. However, the drawback of this solution is that imaging needs to be treated as coherent, which is known to lead to ringing and reduced spatial resolution in case of non-interferometric microscopy [48]. Our calculations predict this effect also for QPI. Hence, confocal QPI provides slightly better performance than common-path QPI when imaging small objects at the diffraction limit.

Improved spatial resolution, absence of halo and ringing as well as suppression of out-of-plane scattering are advantages that make confocal QPI well suited for the label-free imaging of biological specimens. As a demonstration of this application potential, we next compare QPI performance for label-free imaging of tissue sections. To this end, we first built a 2D refractive index model of a thin section of breast tissue that accurately reproduces shape and distribution of subcellular components with high resolution (Fig. 6a). As a template we employed an ultrastructural image of breast that was acquired with electron microscopy and thus affords nanoscale spatial resolution [49]. We labeled and assigned typical values for the refractive index to the individual cell components including cytoplasm (*n* = 1.375), nucleus (1.36), nucleolus (1.38), mitochondria (1.41) [50], lipid droplets (1.48) [51], golgi apparatus, endoplasmic reticulum and microvilli (membranes, 1.46) [52] and assume extracellular medium (1.35). Fig. 6b shows the calculated phase images. The spatial resolution advantage of confocal microscopy is apparent as well as the absence of ringing, producing a visibly cleaner image in comparison to WF methods. This interesting observation indicates that confocal QPI might have an advantage over wide-field QPI in resolving malignancy-induced morphological alterations, especially those located at the sub-microscopic level, which could help to reveal indicators of malignancies more reliably and should be explored further.

To illustrate degradation of image contrast by halo, we built a low-resolution model of the breast tissue section from data shown in Fig. 1f. This was done by linearly converting optical phase contrast to topography. The resulting model is a binary ($n$ = 1.38, 1.40) 3D refractive index map of 2.5 μm thickness, as illustrated by ($x,y$) and ($x,z$) cross sections in Fig. 6c. We observe that the halo of common-path WF QPI yields similar brightness between epithelial and extracellular regions as well as bright patches within the epithelium (Fig. 6d). This effect confounds the analysis of global tissue architecture. Microscopic features such as size and shape cell nuclei seem to be correctly reproduced despite halo, although it is more difficult to locate them because of a lack of contrast. Confocal QPI and traditional WF QPI yield the correct contrast, which could benefit accuracy of manual inspections as well as automated diagnosis by machine learning algorithms. We note that halo in common-path WF QPI is mitigated in the limit of zero condenser NA. However, mechanical limitations of the microscope ($\text{NA}_{\text{cond}} \geq 0.09$) and shot noise (light throughput with halogen lamps) set a lower practical limit to $\text{NA}_{\text{cond}}$. Here we assumed $\text{NA}_{\text{cond}} = 0.045$ which is in the typical range of common-path WF QPI.

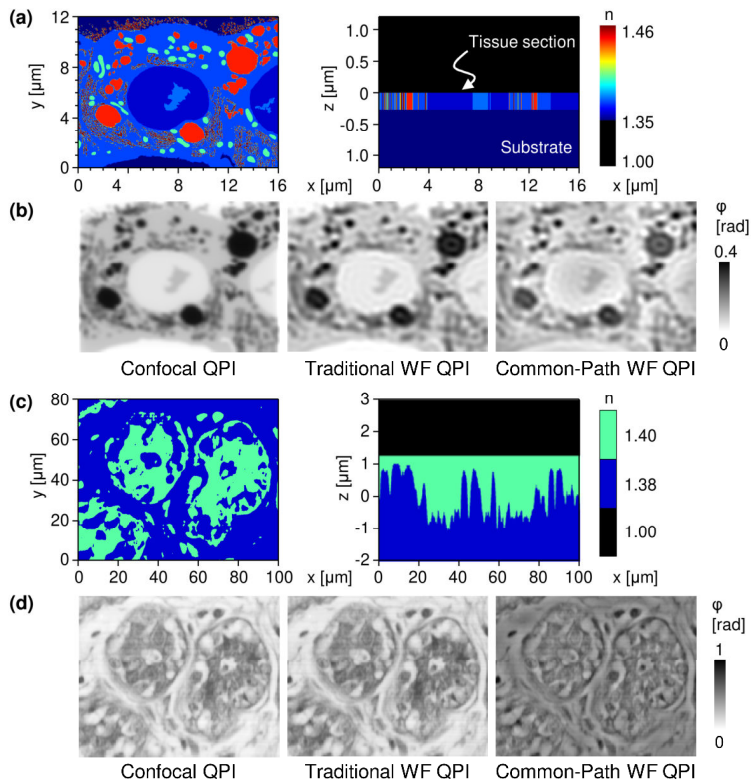

Fig. 6: Calculated phase contrast for numerical tissue section models. (a) High-resolution 2D refractive index model adopted from ultrastructural electron microscopy data on breast. (b) Calculated phase images illustrate improved spatial resolution and absence of ringing artifacts with confocal QPI ($\text{NA}_{\text{cond}} = 0.09$ and $\text{NA}_{\text{obj}} = 0.8$). (c) Large

area, low resolution model of breast tissue constructed from the data in Fig 1. (d) Calculated phase images illustrate the effect of halo on image contrast that appears with common-path wide-field QPI ($\mathrm{NA_{cond}} = 0.045$ and $\mathrm{NA_{obj}} = 0.4$).

## 5. Discussion and Conclusions

We have demonstrated high-resolution label-free imaging of tissue sections with confocal QPI as a means to provide morphological detail regarding disease state. This study presents a significant step for QPI towards digital pathology by providing two advances. The first major result is that QPI can reproduce the spatial level of detail and shading (in greyscale) with similar quality as compared to imaging of H&E stained slides, the gold standard in histopathology, because all spatial frequencies are captured. This is an advance over previous QPI studies based on WF methods that produced phase images of tissue sections with comparatively lower contrast because of the high-pass filtering effect of halo. Automatic segmentation and disease detection has thus mainly been based on the analysis of spatially localized features such as scattering length which is less affected by the high-pass filtering effect. Our results show that confocal QPI can measure the refractive index of tissue sections in absolute terms, which makes manual inspection more reliable and could be used in future for more accurate automated segmentation. As it is typical for label-free QPI imaging, also with confocal QPI the specificity and hence contrast between different tissue components was found to be reduced compared to H&E. However, this loss in specificity was compensated to some degree by the high spatial detail, still allowing nuclei to be discerned in individual glands, for example. As a way to increase specificity, confocal QPI is in principle amenable to co-registered phase and fluorescence imaging. To this end, one could employ sample illumination at two wavelengths simultaneously, one for the fluorescence channel and one for the phase channel, and make use of the spectral detector units, that commercial systems come equipped with, to separate both signals.

Further, the results of this study demonstrate for the first time the value of FDTD calculations to obtain an accurate prediction of QPI performance for tissue imaging. First, generic QPI system can be modeled by rigorously solving Maxwell's equations. Particularly, it is not required to develop specific analytical models to describe image contrast and thus provides for a more flexible simulation platform. Secondly, arbitrary samples can be simulated where multiple scattering and resonance effects are taken into account. This offers the opportunity to use high-resolution electron microscopy data of tissue sections to build a numerical tissue phantom, as it was demonstrated here with two simple models. Specifically for the case of confocal QPI, calculations predicted improved image quality in terms of spatial resolution and contrast when compared to traditional WF methods with low condenser settings. Further work could address the question how multiple scattering and speckle affect phase imaging in case of relatively thick and heterogeneous samples such as the standard histology slides.

While confocal microscopy is a widely spread technique, confocal QPI is not trivial: careful alignment of the optical setup, phase instabilities caused by environmental influences and slow measurements are typically associated with interferometry-based confocal phase imaging. Our implementation based on SOH addresses these challenges by providing rapid, alignment-free QPI on a beam-scanning commercial confocal instrument with a Mirau interference objective. Confocal QPI holds promise to simplify histopathologic analysis by removing the need for sample staining. Synergy with IR microscopy could lead to transformative imaging modalities where high-resolution morphology data provided by confocal QPI is combined with highly specific but low-resolution recognition of cell type and disease state provided by IR microscopy.

The practicality of confocal QPI for application in histopathology is determined by a set of factors. First, imaging speed in confocal QPI is determined by the speed of the galvo scanners as it is the case in non-QPI confocal microscopy. Our approach for QPI is based on holography and thus requires oversampling, which increases imaging times by factor of ~6. Current imaging times are in the range of 14 minutes for complete acquisition of 1.5 mm diameter tissue cores, which is about a magnitude slower than typical WF QPI systems. Higher imaging speed may be achieved by employing faster galvo scanners or resonant scanners, both available commercially. Alternatively, line scanning confocal microscopes are in principle amenable to our implementation of confocal QPI based on SOH, which could improve imaging speed significantly. Secondly, variation in refractive index among the different subcellular constituents is weak and thus a sufficiently high signal-to-noise (SNR) ratio is needed. Confocal QPI is in principle capable of reaching high SNR owing to laser illumination and light detection with a photo detector, significantly reducing shot noise. SNR of the presented system was determined to be 13 mrad RMS [44], being mainly limited by phase stability, which is already sufficient to resolve morphology in tissue sections of standard (5 µm) thickness. We anticipate further improvements in terms of spatial resolution by using higher NA objectives, higher SNR by improved isolation of the system from environmental influences, and improved imaging times. The presented FDTD method can be utilized to model the impact of such modifications on the image quality and be used to guide instrument development.

**Acknowledgements**

M. S. acknowledges support by the European Union's Horizon 2020 research and innovation programme under the Marie Sklodowska-Curie grant agreement No 655888. T.P.W. was supported by a Beckman Institute Postdoctoral Fellowship from the Beckman Institute for Advanced Science and Technology, University of Illinois at Urbana–Champaign.

**Competing interests:** M. S. and P. S. C. are authors of US patent 9,213,313.

**ORCID**
MS 0000-0003-3514-3127
SG 0000-0001-9571-0336
MGD 0000-0003-2593-8268
RB 0000-0001-7360-994X.

19. P. Wang, R. Bista, R. Bhargava, R. E. Brand, and Y. Liu, "Spatial-domain low-coherence quantitative phase microscopy for cancer diagnosis," Opt. Lett. **35**, 2840 (2010).
20. H. Majeed, M. E. Kandel, K. Han, Z. Luo, V. Macias, K. Tangella, A. Balla, and G. Popescu, "Breast cancer diagnosis using spatial light interference microscopy," J. Biomed. Opt. **20**, 111210 (2015).
21. C. Maurer, A. Jesacher, S. Bernet, and M. Ritsch-Marte, "Phase contrast microscopy with full numerical aperture illumination," Opt. Express **16**, 19821 (2008).
22. C. Edwards, B. Bhaduri, T. Nguyen, B. G. Griffin, H. Pham, T. Kim, G. Popescu, and L. L. Goddard, "Effects of spatial coherence in diffraction phase microscopy," Opt. Express **22**, 5133 (2014).
23. T. H. Nguyen, C. Edwards, L. L. Goddard, and G. Popescu, "Quantitative phase imaging with partially coherent illumination," Opt. Lett. **39**, 5511 (2014).
24. P. Hosseini, R. Zhou, Y.-H. Kim, C. Peres, A. Diaspro, C. Kuang, Z. Yaqoob, and P. T. C. So, "Pushing phase and amplitude sensitivity limits in interferometric microscopy," Opt. Lett. **41**, 1656 (2016).
25. Z. Wang, L. Millet, M. Mir, H. Ding, S. Unarunotai, J. Rogers, M. U. Gillette, and G. Popescu, "Spatial light interference microscopy (SLIM)," 11 (2011).
26. Y. Choi, T. D. Yang, K. J. Lee, and W. Choi, "Full-field and single-shot quantitative phase microscopy using dynamic speckle illumination," Opt. Lett. **36**, 2465 (2011).
27. B. Bhaduri, H. Pham, M. Mir, and G. Popescu, "Diffraction phase microscopy with white light," Opt. Lett. **37**, 1094 (2012).
28. G. E. Sommargren, "Optical heterodyne profilometry," Appl. Opt. **20**, 610 (1981).
29. H. J. Matthews, D. K. Hamilton, and C. J. R. Sheppard, "Surface profiling by phase-locked interferometry," Appl. Opt. **25**, 2372 (1986).
30. G. Barbastathis, M. Balberg, and D. J. Brady, "Confocal microscopy with a volume holographic filter," Opt. Lett. **24**, 811 (1999).
31. J. V. Knuuttila, P. T. Tikka, and M. M. Salomaa, "Scanning Michelson interferometer for imaging surface acoustic wave fields," Opt. Lett. **25**, 613 (2000).
32. J. E. Graebner, "Optical scanning interferometer for dynamic imaging of high-frequency surface motion," in *2000 IEEE Ultrasonics Symposium. Proceedings. An International Symposium (Cat. No.00CH37121)* (IEEE, 2000), Vol. 1, pp. 733–736.
33. G. G. Fattinger and P. T. Tikka, "Modified Mach–Zender laser interferometer for probing bulk acoustic waves," Appl. Phys. Lett. **79**, 290–292 (2001).
34. N. Lue, W. Choi, K. Badizadegan, R. R. Dasari, M. S. Feld, and G. Popescu, "Confocal diffraction phase microscopy of live cells," Opt. Lett. **33**, 2074 (2008).
35. A. S. Goy and D. Psaltis, "Digital confocal microscope," Opt. Express **20**, 22720 (2012).
36. A. S. Goy, M. Unser, and D. Psaltis, "Multiple contrast metrics from the measurements of a digital confocal microscope," Biomed. Opt. Express **4**, 1091 (2013).
37. C. Joo, T. Akkin, B. Cense, B. H. Park, and J. F. de Boer, "Spectral-domain optical coherence phase microscopy for quantitative phase-contrast imaging," Opt. Lett. **30**, 2131 (2005).
38. M. A. Choma, A. K. Ellerbee, C. Yang, T. L. Creazzo, and J. A. Izatt, "Spectral-domain phase microscopy," Opt. Lett. **30**, 1162 (2005).
39. M. Schnell, P. S. Carney, and R. Hillenbrand, "Synthetic optical holography for rapid nanoimaging," Nat. Commun. **5**, 3499 (2014).
40. M. Schnell, M. J. Perez-Roldan, P. S. Carney, and R. Hillenbrand, "Quantitative confocal phase imaging by synthetic optical holography," Opt. Express **22**, 15267 (2014).
41. C. Liu, S. Marchesini, and M. K. Kim, "Quantitative phase-contrast confocal microscope," Opt. Express **22**, 17830 (2014).
42. C. Liu, S. Knitter, Z. Cong, I. Sencan, H. Cao, and M. A. Choma, "High-speed line-field confocal holographic microscope for quantitative phase imaging," Opt. Express **24**, 9251 (2016).
17